\documentclass[12pt]{article}
\usepackage{amssymb}
\usepackage{amsmath}
\usepackage{mathrsfs}

\begin{document}
\thispagestyle{empty}

\begin{flushright}
 KEK-TH-1259
\end{flushright}

\vspace{10mm}
\begin{center}
{\Large \bf Generalized Conformal Symmetry  and Recovery
of $SO(8)$ 
in Multiple M2 and D2 Branes}

\vspace{10mm}

Yoshinori Honma$^{a}$\footnote{yhonma@post.kek.jp},
Satoshi Iso$^{a}$\footnote{satoshi.iso@kek.jp},
Yoske Sumitomo$^{a}$\footnote{sumitomo@post.kek.jp},
 Hiroshi Umetsu$^b$\footnote{umetsu@kushiro-ct.ac.jp}\\
and  Sen Zhang$^{a}$\footnote{zhangsen@post.kek.jp}

\vspace{10mm}
 $^a${\it Institute of Particle and Nuclear Studies, \\ High Energy Accelerator Research Organization(KEK)  \\
and \\
 Department of Particles and Nuclear Physics, \\
 The Graduate University for Advanced Studies (SOKENDAI), 
 \\
Oho 1-1, Tsukuba, Ibaraki 305-0801, Japan} 
\vspace{5mm} \\
$^b${\it Department of General Education, Kushiro National College
of Technology \\
Otanoshike-Nishi 2-32-1, Kushiro 084-0916, Japan}
\end{center}

\vspace{20mm}
\ \ 

\begin{abstract}
We investigate  conformal symmetries  of the Aharony - Bergman -
 Jafferis - Maldacena (ABJM) theory 
for multiple M2 branes  and the Lorentzian
Bagger - Lambert - Gustavsson  (L-BLG) theory which can be
obtained by taking a scaling limit 
$k ( \gg N) \rightarrow \infty$ of the ABJM theory. 
The conformal symmetry is maintained in the L-BLG
by considering general space-time varying solutions
to the constraint equations. 
The dual geometry is reduced to  $d=10$ 
$AdS_4 \times {\bf CP}^3$ in the scaling 
limit and has the same conformal symmetry.
The curvature radius $R$ 
satisfies $l_p^{(11)} \ll l_p^{(10)} \ll R \ll l_s$ 
($l_p^{(d)}$ and $l_s$ are the $d$-dimensional Planck
lengths and the string scale), 
and the theory is in a region where an $\alpha'$ expansion
is not valid.
We also study how the $SO(8)$ covariance is recovered
in the $AdS_4 \times {\bf CP}^3$ geometry
by taking the scaling limit. 
\end{abstract}

\newpage
\setcounter{page}{1}
\setcounter{footnote}{0}

\baselineskip 6mm
\section{Introduction}
\setcounter{equation}{0}
Bagger and Lambert, and Gustavsson
discovered the ${\cal N}=8$ superconformal
$(2+1)$-dimensional field theories with a $SO(8)$
global symmetries by exploiting the 3-algebraic
structures \cite{Bagger:2007jr,Gustavsson:2007vu} 
and an effective theory of multiple M2 branes
was proposed, which was
based on the 3-algebra with a Lorentzian signature
\cite{Gomis:2008uv,Benvenuti:2008bt,Ho:2008ei}
(the L-BLG theory).
Another very interesting proposal  for multiple
M2 branes  was also made recently  by 
Aharony, Bergman, Jafferis and
Maldacena (the ABJM theory) \cite{Aharony:2008ug}.

An earlier proposal of L-BLG
\cite{Gomis:2008uv,Benvenuti:2008bt,Ho:2008ei}
has the desired symmetries, $SO(8)$ and superconformal
symmetries but there are several problems.
First, because of the Lorentzian signature,
the model contains  fields $X_0^I$ and $X_{-1}^I$ which
may endanger the unitarity of the theory.
The fields $X_{-1}^I$ (and
the fermionic partner $\Psi_{-1}$) are, however, contained
in the action only linearly and can be
 integrated out to give the following constraints on 
the `conjugate' fields $X_0^I$ and $\Psi_0$:
\begin{equation}
\partial^2 X_0^I =0, \ \ \Gamma^\mu \partial_\mu \Psi_0 =0. 
\label{constraints}
\end{equation}
Hence the would-be ghost modes can be removed from the
propagating degrees of freedom.\footnote{There are  attempts
to kill the ghost fields by gauging a shift symmetry 
 \cite{Bandres:2008kj,Gomis:2008be}.}
If we take a constant solution $X_0^I=v^I$ to the 
constraint equation, the L-BLG
 theories are 
reduced to the action of the $N$ D2 branes
in flat space \cite{Mukhi:2008ux,Ho:2008ei,Honma:2008un,Ezhuthachan:2008ch}.
The specific choice of the solutions breaks the conformal
invariance and the $SO(8)$ to $SO(7)$
and it has been  suspected that the L-BLG theory 
is nothing more than a theory of  D2 branes in $d=10$ spacetime. 
This is another problem of the L-BLG theory.

In this paper we would like to emphasize 
that the constraint equations 
(\ref{constraints}) should be more carefully treated
as we did in \cite{Honma:2008un,Honma:2008jd} 
and show that the interpretation of the L-BLG as the 
ordinary D2 branes is not
appropriate (see also a recent work \cite{Verlinde:2008di}). 
In  \cite{Honma:2008un}, 
we revisited the constraint equation
and considered  general
spacetime dependent solutions and the theory around 
it
\footnote{In \cite{Honma:2008un}, we have also studied   
the constraint equations of the  mass-deformed theory 
\cite{Gomis:2008cv,Hosomichi:2008qk}.
In this case, the constraint equation is modified to
$ (\partial^2 + \mu^2) X_0^I =0 $
and there are no constant solutions. 
We studied the theory around a background of the
spacetime dependent solution 
$ X_0^I = \exp(\mu x) \ \delta^I_{10} $
and called such field theories Janus field theories.
It was also shown in \cite{Blau:2008bm}  
 that the spacetime dependent coupling in the massless theory can be
 reinterpreted as a coordinate-dependent mass term with a constant
 coupling.}.
By considering such space-time dependent solutions,
the theory is shown to have a {\it generalized conformal symmetry}
as well as the manifest $ SO(8)$ invariance
\cite{Honma:2008jd}. The purpose of the present paper
is to investigate these symmetries extensively both
in the field theories and in the gravity duals.

Another very interesting proposal  for multiple
M2 branes  was made by 
Aharony et.al. \cite{Aharony:2008ug}. They generalized the
superconformal Chern-Simons matter theories 
\cite{Schwarz:2004yj,Gaiotto:2007qi} to the ${\cal N}=6$
superconformal $U(N) \times U(N)$ theories.
The levels of the Chern-Simons gauge fields
are $(k,-k)$ and the theory is conjectured to
 describe the low energy limit of $N$ M2-branes
 probing  ${\bf C}^4/{\bf Z}_k$. Hence at large $N$, it is 
dual to the M-theory on 
$AdS_4 \times S^7/{\bf Z}_k$.
In the formulation the 3-algebraic structure does not seem to play 
any role, but recently Bagger and Lambert showed
that if the condition of the antisymmetry of the 
structure constant is relaxed, the ABJM theory
can be written in terms of the new 3-algebra \cite{Bagger:2008se}.
There are also interesting works of the ABJM theory for the relation to a M5-brane
through the Basu-Harvey equations
\cite{Terashima:2008sy,Gomis:2008vc,Hanaki:2008cu}, ${\cal N}=6$ Chern-Simons theory with the
other gauge groups \cite{Hosomichi:2008jb,Schnabl:2008wj} and the
${\cal N} \leq 8$ superconformal theories from the 3-dimensional gauged
supergravity models \cite{embedding}.
For other related works of the ABJM theory, see \cite{others}.

Now we have two different proposals for the formulation
of multiple M2 branes, L-BLG and ABJM, 
and it is important to 
understand the relation between them, especially
how the conformal symmetries and $SO(8)$ invariance
are realized in these theories.
In  \cite{Honma:2008jd}, we have explicitly shown
that the L-BLG theory can be obtained by 
taking the following scaling limit 
of the ABJM theory (fermions are omitted here):
\begin{align}
 B_{\mu} &\rightarrow \lambda B_{\mu}, \nonumber \\
 X_0^I &\rightarrow \lambda^{-1} X_0^I, \nonumber \\
 k &\rightarrow \lambda^{-1} k,
\end{align}
where  $B_\mu$ is an axial combination of the two
gauge fields $B_\mu=(A^{(L)}_\mu - A_\mu^{(R)})/2$ and
$X_0^I$ are  trace components of the bifundamental matter
fields.
The other bosonic fields are kept fixed
and  then  take the limit:
\begin{align}
\lambda \rightarrow 0.
\end{align}
The gauge group $SU(N) \times SU(N)$ of the ABJM theory
is  reduced to the semi-direct sum of $SU(N)$ and 
translations a l\'a
In\"on\"u-Wigner
contraction \cite{Inonu:1953sp}.
Taking the scaling limit, the action of the ABJM theory
is reduced to the action of the L-BLG theory.  
Furthermore,  the 
same constraint equations (\ref{constraints}) can be obtained by 
requiring finiteness of the action 
in the  $\lambda \rightarrow 0$ limit.
We emphasize here that the scaling limit is taken before
taking the large $N$ limit. Hence the 't Hooft coupling 
$N/k$ vanishes
in the scaling limit from ABJM to L-BLG.

The M2 branes described by the ABJM theory
 has conformal symmetry.
The scaling limit mentioned above
corresponds to  locating M2 branes
far from the origin of the ${\bf Z}_k$ orbifold
as well as taking $k (\gg N)\rightarrow \infty$.
Since the coupling of the 
scaled theory is promoted to an 
$SO(8)$ vector $X_0^I(x)$, we showed in \cite{Honma:2008jd} that
the scaled theory of L-BLG has an enhanced symmetries, i.e.
{\it generalized conformal symmetry} and 
$SO(8)$  invariance. 
These symmetries are not expected to exist
 in the effective theory of the ordinary D2 branes.
This generalized conformal
symmetry is essentially the same as that
proposed  by Jevicki, Kazama 
and Yoneya \cite{Jevicki} 10 years ago for 
general Dp-branes.

In this paper we  further investigate  the 
conformal symmetry and recovery of $SO(8)$
invariance 
in the ABJM and L-BLG theories. 
In section 2, we first analyze the conformal invariance
of the ABJM theory, in particular the invariance under
the special conformal transformations.
Since the scaling limit is compatible with the
conformal invariance, the L-BLG theory also has the
same conformal invariance.
We also show  the constraint 
equations (\ref{constraints})
are compatible with the conformal symmetries.
It should be emphasized that the 
conformal invariance can be preserved 
only when we consider a  set
of spacetime dependent solutions to the constraint 
equations, and a specific choice to the equations
generally breaks the conformal invariance.

In section 3, we discuss the conformal symmetry
and the recovery of $SO(8)$ 
in the gravity dual. 
In \cite{Aharony:2008ug}, the dual geometry of the
ABJM theory is conjectured to be
$AdS_4 \times S^7/{\bf Z}_k$ where the 
$S^7$ is considered as $U(1)$ Hopf fibration on ${\bf CP}^3$
and the $U(1)$ direction is orbifolded.
In the scaling limit of ABJM to L-BLG, $k$ is taken to infinity
and the ${\bf Z}_k$ identification looks like a circle 
identification of the $d=11$ theory. 
In this reduction to $d=10$,
the dilaton field takes a constant value
and the reduced $d=10$ geometry is given by
$AdS_4 \times {\bf CP}^3$.
In the original discussion of ABJM, 
 the 't Hooft coupling $N/k$ 
is kept fixed and hence the radius of curvature of $AdS_4$
is finite in the string units.
However in our scaling limit to L-BLG, $k$ is taken to 
 infinity before taking the large $N$
 and the radius becomes
much smaller than the string scale:
\begin{align}
\left( \frac{R}{l_s} \right)^2 =
R_{str}^2 \sim \sqrt{\frac{N}{k}} \rightarrow 0.
\end{align}
On the other hand,
comparing $R$ with the $d=10$ Planck length,
the ratio is given by 
\begin{align}
\left( \frac{R}{l_p^{(10)}} \right)^2 
\sim k^{1/8} N^{3/8} \rightarrow \infty
\end{align}
and the type IIA supergravity approximation itself is good.
Hence the reduced geometry of ABJM in the scaling limit
$k \rightarrow \infty$
to L-BLG can be described by 
$AdS_4 \times {\bf CP}^3$,
but it cannot be considered as a low energy 
approximation of type IIA superstring.
The scaled theory (L-BLG) may be  more appropriately
interpreted as M2 branes in $d=10$ that is 
dimensionally reduced from the $d=11$ supergravity.

In Appendix \ref{sec:u1-part-abjm}, we discuss the effect 
of $U(1)$ gauge field in the scaling limit of the 
$U(N) \times U(N)$ ABJM theories.
In Appendix \ref{appendixB}, the recovery of $SO(8)$ 
in ${\bf C}^4/U(1)$ is discussed.
In Appendix \ref{sec:ordin-reduct-m2}, we review the ordinary
reduction from $d=11$  M2 branes to $d=10$ D2 branes.


\section{Conformal Symmetry of ABJM and L-BLG}
\setcounter{equation}{0}
\subsection{Conformal invariance of ABJM}
The ABJM 
theory of $d=3$ ${\cal N}=6$ superconformal theory
is proposed  as a dual field theory of 
 M-theory on $AdS_4 \times S^7/{\bf Z}_k$.
As shown in \cite{Bandres:2008ry}, the ABJM theory is 
invariant under the superconformal transformations. 
Here we study the invariance of the ABJM theory 
under the conformal transformations, in particular
 the special conformal transformations.

The action of the ABJM theory is given by
\begin{align}
 {\cal S} &= \int d^3 x \ \textrm{tr} [- (D_{\mu}Y_A)^{\dagger}D^{\mu} Y^A
   + i \psi^{\dagger}_A
  \Gamma^{\mu} D_{\mu} \psi^A  ] + {\cal S}_{CS} - {\cal S}_{V_f} - {\cal S}_{V_b}
\end{align}
where
\begin{align}
 &{\cal S}_{CS}  = \int d^3 x \ 2K \epsilon^{\mu\nu\lambda}
  \textrm{tr} [A^{(L)}_{\mu}\partial_{\nu} A^{(L)}_{\lambda} + \frac{2i}{3}
  A^{(L)}_{\mu}A^{(L)}_{\nu}A^{(L)}_{\lambda} 
  - A^{(R)}_{\mu}\partial_{\nu}
  A^{(R)}_{\lambda} - \frac{2i}{3}
  A^{(R)}_{\mu} A^{(R)}_{\nu} A^{(R)}_{\lambda}], \notag\\
 &S_{V_b} = -\frac{1}{48 K^2} \int d^3 x \ \textrm{tr}[
  Y^A Y^{\dagger}_A Y^B Y_B^{\dagger} Y^C Y^{\dagger}_C
  + Y^{\dagger}_A Y^A Y^{\dagger}_B Y^B Y_C^{\dagger} Y^C \nonumber \\
   & \ \ \ \ \ \ \ \ \ \ \ \ \ \ \ \ \ \ \ \ \ \ \ \ \ \ \ \ \ \ \ +4 Y^A Y^{\dagger}_B Y^C Y^{\dagger}_A Y^B Y_C^{\dagger}
   - 6 Y^A Y^{\dagger}_B Y^B Y^{\dagger}_A Y^C Y_C^{\dagger}], \notag\\
 &S_{V_f} = \frac{i}{4K} \int d^3 x \textrm{tr} [Y^{\dagger}_A Y^A
  \psi^{B\dagger} \psi_B - Y^A Y^{\dagger}_A \psi_B \psi^{B\dagger} + 2
  Y^A Y^{\dagger}_B \psi_A \psi^{B\dagger} - 2 Y^{\dagger}_A Y^B
  \psi^{A\dagger} \psi_B \nonumber \\ 
   & \ \ \ \ \ \ \ \ \ \ \ \ \ \ \ \ \ \ \ \ \ \ \ \ \ + \epsilon^{ABCD} Y^{\dagger}_A \psi_B Y^{\dagger}_C \psi_D -
  \epsilon_{ABCD} Y^A \psi^{B\dagger} Y^C \psi^{D\dagger}]\notag
\end{align}
and $A = 1,2,3,4$.
We used the notation  of \cite{Benna:2008zy} and $K = k/8 \pi$.

It is a  $U(N) \times U(N)$ or $SU(N) \times SU(N)$ gauge theory.
The other choices of gauge groups are possible but here we consider these
two types. The actions of the gauge fields are given by the Chern-Simons
action with coefficients $k$ and $-k$.
Matter fields $Y^A$ and $\psi^A$ are in the bifundamental representation
and  the covariant derivative is defined by
\begin{align}
 D_{\mu} Y = \partial_{\mu} Y + iA^{(L)}_{\mu} Y - iY A^{(R)}_{\mu}.
\end{align}
The action is invariant under ${\cal N}=6$ superconformal transformations.
In the following we check the explicit invariance 
under the conformal transformations. 

First it is obvious that 
the action is invariant under the dilatation.
Dilatation is defined by
$ x \rightarrow e^\epsilon x $
and simultaneously 
we transform each field by multiplying $e^{-n \epsilon}$
where $n$ is the conformal weight.
The scalars $Y^A$, fermions $\psi^A$ and the gauge fields
$A_\mu$ have weights $1/2, 1, 1$ respectively.

A little more nontrivial transformation is 
a special conformal transformation.
It is given by
\begin{equation}
\delta x^\mu = 2 \epsilon \cdot x x^\mu - \epsilon^\mu x^2.
\label{sct}
\end{equation}
If we write  the infinitesimal transformation for each field $Y(x)$
as $\delta Y(x)=Y'(x')-Y(x)$, they are given by
\begin{align}
 \delta Y^A(x) &= -\epsilon \cdot x Y^A(x), \notag\\
 \delta A^{(L,R)}_\mu(x) &= -2 \epsilon \cdot x A^{(L,R)}_\mu(x) -
 2(x\cdot A^{(L,R)} \epsilon_\mu - \epsilon \cdot A^{(L,R)} x_\mu ), \notag\\
 \delta \psi^A(x) &= -2 \epsilon \cdot x \psi^A(x) -
 \epsilon_{\mu\nu\lambda}\epsilon^\nu x^\lambda \Gamma^\mu \psi^A(x)
 \label{sctZ}.
\end{align}
These transformations can be understood as follows.
They look like the general coordinate transformations,
but  are different since the theory is restricted to live in 
the flat space-time with a fixed metric
and the change of the metric
under the general coordinate transformations
must be compensated by the transformations of the fields.
The first terms in each transformation reflect the 
conformal weight of each field.
The second term in the transformation of the fermion
is the local Lorentz transformation which pulls back the 
flat local Lorentz frame (where we use $\Gamma^{012}\psi = \psi$).
The transformation for the gauge field $A_{\mu}$ is nothing 
but the general coordinate transformation with the 
transformation parameter (\ref{sct}).

The action is invariant under the above special conformal transformations.
In order to see it, the following transformation rules
are useful: 
\begin{align}
d^3 x  \rightarrow \ & e^{6\epsilon \cdot x} d^3 x, 
 \notag\\
\partial_\mu \rightarrow \ & e^{-2\epsilon \cdot x} [
 \partial_\mu -2(\epsilon_\mu x^\nu\partial_\nu -x_\mu \epsilon^\nu
 \partial_\nu) ],
 \notag\\
D_\mu Y \rightarrow \ & e^{-3\epsilon \cdot x} \left[ D_\mu Y -\{ Y+2x^\nu \partial_\nu Y+2i(x \cdot A^{(L)}Y-Y x \cdot A^{(R)})\} \epsilon_\mu \right. \nonumber \\
 & \left. \ \ \ \ \ \ \ \ \ \ \ \ \ \ \ \ \ +\{ 2\epsilon^\nu\partial_\nu Y+2i(\epsilon \cdot A^{(L)}Y-Y \epsilon \cdot A^{(R)})\} x_\mu \right], \notag\\
 F_{\mu \nu} \rightarrow  & e^{-4\epsilon \cdot x} \left[
 F_{\mu \nu}-2(\epsilon_\nu x^\rho F_{\mu \rho}-\epsilon_\mu x^\rho
 F_{\nu \rho})+2(x_\nu \epsilon^\rho F_{\mu \rho}-x_\mu \epsilon^\rho
 F_{\nu \rho}) \right]
 \label{volume}.
\end{align}
Though $\epsilon$ is an infinitesimal parameter, we write the overall
factors as $e^{-2n \epsilon \cdot x}$ for convenience.  
They are cancelled 
in the action because $n$ is  the conformal weight 
of each field and coordinates.

Here let us 
check the invariance of  the Chern-Simons term as an example.
First the derivative part transforms as 
\begin{align}
&\epsilon^{\mu\nu\lambda} 
\textrm{tr} F_{\mu \nu} A_{\sigma}\notag\\
 &\rightarrow 
 \epsilon^{\mu\nu\lambda} e^{-6\epsilon \cdot x} \textrm{tr}
 [ F_{\mu \nu} A_{\lambda} + 
 4(\epsilon_\mu x^\rho - x_\mu \epsilon^\rho ) A_\lambda F_{\nu \rho}
 - 2 F_{\mu \nu} (x\cdot A \epsilon_\lambda - \epsilon \cdot A x_\lambda ) ].
\end{align}
 The pre-factor $e^{-6\epsilon \cdot x}$ is cancelled with 
 the transformation of $d^3 x$ in (\ref{volume}).
 The rest vanishes because
\begin{align}
& \epsilon^{\mu\nu\lambda} \textrm{tr}
  [  2(\epsilon_\mu x^\rho - x_\mu \epsilon^\rho ) A_\lambda F_{\nu \rho}
 -  F_{\mu \nu} (x\cdot A \epsilon_\lambda - \epsilon \cdot A x_\lambda ) ] 
 \nonumber \\
& = \epsilon^{\mu\nu\lambda} \textrm{tr}
[2\epsilon_\mu^{\ \rho \alpha} f_\alpha F_{\nu \rho}A_\lambda
-\epsilon^{ \ \rho \alpha}_\lambda f_\alpha F_{\mu \nu} A_\rho ]
= 0.
\end{align}
 In the second line we have defined 
 $f^\alpha = \epsilon^{\mu \nu \alpha}x_\mu \epsilon_\nu.$
 Similarly the invariance of the term 
 $\epsilon^{\mu \nu \lambda}A_\mu A_\nu A_\lambda$
 can be shown by noting that the gauge field transforms as
 \begin{equation}
 A_\mu \rightarrow e^{-2\epsilon \cdot x} 
  (A_\mu + 2 \epsilon_{\mu \alpha \beta} f^\alpha A^\beta).
 \end{equation}
 Hence the Chern-Simons terms are invariant under the special 
 conformal transformation.  Though we have checked it explicitly,
 the invariance  can be  naturally understood
  because the Chern-Simons term is independent
 of the metric if it is defined in a curved background space-time.
  
The other terms in the action are also 
straightforwardly shown to be invariant under the 
special conformal transformations.
\subsection{ABJM to L-BLG}
As shown in \cite{Honma:2008jd}, 
the L-BLG 
theory is obtained by taking a scaling limit of the ABJM theory
with a gauge group $SU(N) \times SU(N)$.
In the gauge theory with $U(N) \times U(N)$
there is a subtlety in the scaling of the $U(1)$ part.
We will discuss the issue in the Appendix \ref{sec:u1-part-abjm} and here restrict 
the discussions to the $SU(N) \times SU(N)$ case.
 
The scaling is given as follows:
\begin{align}
 B_{\mu} &\rightarrow \lambda B_{\mu}, \nonumber \\
 X_0^I &\rightarrow \lambda^{-1} X_0^I, \nonumber \\
 \psi_{A0} &\rightarrow \lambda^{-1} \psi_{A0}, \nonumber \\
 k &\rightarrow \lambda^{-1} k 
 \label{scaling}
\end{align}
where  
\begin{equation}
Y^A = X^{2A-1}_0 + i X^{2A}_0 - \hat{X}^{2A} + i \hat{X}^{2A-1}, \ \ \ \  
 B_\mu =\frac{1}{2}(A^{(L)}_\mu - A_\mu^{(R)})
\end{equation}
and  $X_0^I$ and $\psi_{0A}$ are  trace components of the bifundamental matter
fields, and $I=1, \cdots, 8$.
When we take $\lambda \rightarrow 0$ limit and keep the other fields fixed, the action of the ABJM theory is reduced to the action of the L-BLG theory. 
Since the $k \rightarrow \infty$ limit is taken before taking the large $N$,
our scaling corresponds to a vanishing 't Hooft coupling $N/k \rightarrow 0$.
Besides the action, the same constraint equations as those in the L-BLG theory
can be obtained from the ABJM theory:
\begin{equation}
\partial^2 X_0^I =0, \ \ \Gamma^\mu \partial_\mu \Psi_0 =0,
\label{constraints2}
\end{equation}
by requiring  {\it finiteness of the action} in the  $\lambda \rightarrow 0$ limit. 

In the above scaling limit we arrive at 
 the L-BLG theory:
\begin{align}
 {\cal L}_0 = \hbox{Tr}
   \left[ - \frac{1}{2}(\hat{D}_{\mu} \hat{X}^I - B_{\mu} X_0^I)^2
    + \frac{1}{4}(X_0^K)^2 ([\hat{X}^I,\hat{X}^J])^2
    - \frac{1}{2} (X_0^I [\hat{X}^I,\hat{X}^J])^2
    \right. \nonumber \\
     + \frac{i}{2} \bar{\hat{\Psi}} \Gamma^{\mu} \hat{D}_{\mu} \hat{\Psi} +
   i \bar{\Psi}_0 \Gamma^{\mu} B_{\mu} \hat{\Psi} 
      -\frac{1}{2}\bar{\Psi}_0 \hat{X}^I
      [\hat{X}^J,\Gamma_{IJ}\hat{\Psi}]
      + \frac{1}{2}\bar{\hat{\Psi}}X^I_0[\hat{X}^J,\Gamma_{IJ}\hat{\Psi}]
          \nonumber \\
     \left.
    + \frac{1}{2} \epsilon^{\mu\nu\lambda} \hat{F}_{\mu\nu} B_{\lambda}
      - \partial_{\mu} X^I_0\  B_{ \mu} \hat{X}^I
 \right].
 \label{BLaction}
\end{align}
In the original formulation of the L-BLG theory, 
the constraint equations (\ref{constraints2}) are derived by 
integrating the auxiliary fields $X_{-1}^I$ and $\Psi_{-1}$:
\begin{align}
 {\cal L}_{gh} =   (\partial_{\mu} X^I_0)(\partial^{\mu} X^I_{-1}) - i \bar{\Psi}_{-1}
  \Gamma^{\mu} \partial_{\mu} \Psi_0.
  \label{GHaction}
\end{align}
Since the above scaling is compatible with the conformal transformations
discussed in the previous section,  
the action (\ref{BLaction}) is invariant under the conformal transformations (see also \cite{Bandres:2008vf}).
The action for the auxiliary fields (\ref{GHaction}) is also invariant if we define 
the transformations for them as
\begin{align}
\delta X_{-1}^I(x)&=  -\epsilon \cdot x X_{-1}^I(x), \notag\\
\delta \Psi_{-1}(x) &= -2 \epsilon \cdot x \Psi_{-1}(x) -
 \epsilon_{\mu\nu\lambda}\epsilon^\nu x^\lambda \Gamma^\mu \Psi_{-1}(x).
\end{align}

\subsection{Generalized conformal symmetry in  D2 branes}
Now integrate the $B_\mu$ gauge field. 
It has been discussed that if we pick up a specific solution
to the constraint equation (\ref{constraints2}), especially a constant
solution
\begin{equation}
  X_0^I = v  \ \delta^{I,8},  \  \ \ \Psi_0=0,
  \label{constantsolution}
\end{equation}
the L-BLG theory is reduced to the action of the ordinary D2 branes
whose Yang-Mills coupling constant is given by $g_{YM}=v$:
\begin{align}
 {\cal L} = {\rm Tr}\left[- \frac{1}{4 v ^2} \hat{F}_{\mu\nu}^2 -\frac{1}{2} (\hat{D}_{\mu} \hat{X}^A)^2 + \frac{1}{4} v ^2
  [\hat{X}^A,\hat{X}^B]^2 + \frac{i}{2}
  \bar{\hat{\Psi}} \Gamma^{\mu}
  \hat{D}_{\mu} \hat{\Psi} +
  \frac{1}{2} v \bar{\hat{\Psi}}  [\hat{X}^A, \Gamma_{8,A} \hat{\Psi}]
                    \right]
\end{align}
where $A,B=1,\cdots,7$. 
Then $SO(8)$ is spontaneously broken to $SO(7)$ 
because we have specialized the 8-th direction.
The conformal invariance is also broken. 
Though the action is the same as that of the 
D2 branes, we see later that
the interpretation of the L-BLG theory as 
an effective theory of the ordinary
D2 branes is not appropriate since the radius of curvature
is much smaller than the string scale in the gravity dual.

The constraint equations (\ref{constraints2}) have more general 
solutions than (\ref{constantsolution}) which
depend on the spacetime coordinates.
Then  the resulting action becomes a Yang-Mills theory
with a spacetime  dependent coupling  \cite{Honma:2008un}.
As  we have shown \cite{Honma:2008jd},
the action with the spacetime dependent coupling is invariant under
the conformal transformations if we consider a set of spacetime
dependent solutions.
The conformal invariance is discussed 
in more details in  the next section. 

We here consider the simplest spacetime dependent solutions:
\begin{equation}
  X_0^I = v(x) \, \delta^{I,8},\quad \Psi_0=0, \quad \partial^2 v(x) = 0.
\end{equation}
Then 
the L-BLG theory is reduced to the same action as that of the D2 branes
but with a spacetime varying coupling:  
\begin{align}
 {\cal L} = {\rm Tr} \left[ - \frac{1}{4  v(x)  ^2} \hat{F}_{\mu\nu}^2 -\frac{1}{2} (\hat{D}_{\mu} \hat{X}^A)^2 + \frac{1}{4}  v(x)^2
  [\hat{X}^A,\hat{X}^B]^2 \right. \notag\\
 \left. + \frac{i}{2}
  \bar{\hat{\Psi}} \Gamma^{\mu}\hat{D}_{\mu} \hat{\Psi}
 +\frac{1}{2} v(x) \bar{\hat{\Psi}}  [\hat{X}^A, \Gamma_{8,A} \hat{\Psi}]
                    \right].
\label{D2v(x)}
\end{align}
$SO(8)$ symmetry is spontaneously broken to $SO(7)$ as well,
but the action with a varying $v(x)$ has a generalized conformal
symmetry if the coupling transforms as 
\begin{equation}
\delta v(x) =  -(\epsilon \cdot x) \ v(x).
\label{sctv}
\end{equation}
This transformation is originated in the special conformal 
transformation of the scalar field (\ref{sctZ}).
The generalized conformal transformation for Dp branes were first
proposed by Jevicki, Kazama and Yoneya  \cite{Jevicki}.
In the present case, the transformation (\ref{sctv})
is naturally derived  since the coupling constant of the Yang-Mills 
action is determined by the center of mass coordinates $X_0^I(x)$ of 
the M2 branes. 

It is  worth noting that the generalized conformal transformation
(\ref{sctv}) is compatible with the constraint equations (\ref{constraints2})
only when $p=2$.  We will discuss it in the next section.

\subsection{Conformal symmetry and $SO(8)$ invariance of L-BLG}
The space-time dependent coupling  $v(x)$ can be promoted to an
$SO(8)$ vector $X_0^I(x)$
by considering  general solutions to
the constraint equations (\ref{constraints2}) as shown in 
 \cite{Honma:2008un}. 
Then the resultant action after integrating the $B_{\mu}$
gauge field becomes D2 branes effective action with  space-time dependent
couplings in a  vector representation of the $SO(8)$ .  
 In \cite{Honma:2008jd}
 we  showed that if we consider 
space-time dependent solutions the theory has
the {\it generalized conformal symmetry} as well as
the manifest  $SO(8)$ invariance.


In this section we study more details of the generalized conformal
symmetry of the L-BLG theory. 
Especially we show that
the conformal transformations are closed under 
the constraint equations (\ref{constraints2}).  

By integrating the $B_\mu$ gauge field, we get the action
$S=\int d^3 x({\cal L}_{0} + {\cal L}')$:
\begin{align}
{\cal L}_{0} &=
{\rm Tr}\left[  -\frac{1}{2} (\hat{D}_{\mu} P^I)^2 + \frac{1}{4} X_0^2
  [P^I, P^J]^2 + \frac{i}{2}
  \bar{\hat{\Psi}} \Gamma^{\mu}
  \hat{D}_{\mu} \hat{\Psi}  +
  \frac{1}{2}  \bar{\hat{\Psi}}  [P^I, (X_0^J \Gamma_{J}) \Gamma_{I} \hat{\Psi}]
  \right.  \nonumber \\ 
& \hspace{4em}\left.
+\frac{1}{2 (X_0^I)^2} \big(  
\frac{1}{2} \epsilon^{\mu \nu \lambda} \hat{F}_{\nu \lambda} + i \bar{\Psi}_0 \Gamma^\mu \hat{\Psi}
- 2  P_I \partial^\mu X_0^I  \big) ^2 
-\frac{1}{2} \bar{\Psi}_0 \Gamma_{IJ} \hat{\Psi} [P^I,P^J]
                    \right], \notag\\
{\cal L}' &= \frac{1}{X_0^2}  {\rm Tr} \left[
\left( -\bar{\Psi}_0 \Gamma_{I}
(X_0^J \Gamma_{J}) [P^I, \hat{\Psi}] - i \bar{\Psi}_0 \Gamma_{\mu} \hat{D}_\mu \hat{\Psi}
\right)
(X_0^K \hat{X}^K) \right].\label{Janus}
\end{align}
where we have defined a new scalar field $P_{I}$
with  7 degrees of freedom by using the projection 
operator
\begin{equation}
P_{I}(x) = \left( \delta_{IJ} - \frac{X_{0I} X_{0J}}{X_0^2} \right)X^J.
\end{equation}
The $X_0^I(x)$ field is constrained to satisfy $\partial^2 X_0^I=0.$ 
This is a generalization of (\ref{D2v(x)}).
We called this theory a Janus field theory of (M)2-branes since
the coupling constant is varying with the space-time coordinates.

The action of the gauge field is no longer the Chern-Simons action
but we can again show that it is invariant under the 
conformal transformations.
Under the dilatation $x^\mu \rightarrow e^\epsilon x^\mu$, 
each field is multiplied by $e^{-n \epsilon}$ where 
$n$ is the conformal weight. The weights of 
$P,X_0,A_\mu, \Psi, \Psi_0$ are
$1/2, 1/2,1,1,1$ respectively.
The action is evidently invariant.

Special conformal transformation is similarly given by
\begin{equation}
\delta x^\mu = 2 \epsilon \cdot x x^\mu - \epsilon^\mu x^2
\label{sct2}
\end{equation}
and the fields transform as
\begin{align}
\delta P^I(x) &= -\epsilon \cdot x P^I(x), \notag\\
\delta X_0^I(x)&=  -\epsilon \cdot x X_0^I(x), \notag\\
\delta A_\mu(x) &= -2 \epsilon \cdot x A_\mu(x) -
 2(x\cdot A \ \epsilon_\mu - \epsilon \cdot A \ x_\mu ), \notag\\
\delta \hat{\Psi}(x) &= -2 \epsilon \cdot x \hat{\Psi}(x) -
\epsilon_{\mu\nu\lambda}\epsilon^\nu x^\lambda \Gamma^\mu \hat{\Psi}(x),\notag \\
\delta \Psi_0(x) &= -2 \epsilon \cdot x \Psi_0(x) -
 \epsilon_{\mu\nu\lambda}\epsilon^\nu x^\lambda \Gamma^\mu \Psi_0(x).
\end{align}
It is now straightforward to show the invariance of the action.
The Lagrangian is not invariant but changes by  total derivatives.

Finally we need to check that the transformation is closed within
the constraint equations (\ref{constraints2}). 
Namely if the field $X_0^I(x)$ satisfies $\partial_x^2 X_0^I(x)=0$,
the transformed field $X_0^{'I}(x')$ must also satisfy
$\partial_{x'}^2 X_0^{'I}(x') =0$. 
For an infinitesimal special conformal transformation, this is 
equivalent to show $\partial^2 \tilde{\delta} X_0^I(x)=0$
where $\tilde{\delta} X_0^I(x)$ is the 
 transformation at the numerically same point defined by 
\begin{align}
\tilde{\delta}X_0^I(x) = X_0'^I(x) -X_0^I(x) = \delta X_0^I(x) - \delta x^\mu \partial_\mu X_0^I(x),\notag\\
\tilde{\delta} \Psi_0(x) = \Psi_0'(x) -\Psi_0(x) = \delta \Psi_0(x) - \delta x^\mu \partial_\mu \Psi_0(x).
\end{align}

In the following, in order to see the specialty for M2 (or D2)-branes,
we generalize the  special conformal transformation 
to Dp-branes \cite{Jevicki}:
\begin{align}
\tilde{\delta}X_0^I(x)  & = -(3-p)\epsilon \cdot x X_0^I -(2\epsilon \cdot x x^\mu-\epsilon x^2) \partial_\mu X_0^I
\end{align}
It is easy to show 
\begin{align}
\partial^2 (\tilde{\delta}X_0^I(x))
& = 2(p-2)\epsilon^\mu \partial_\mu X_0^I
\end{align}
where we have used the constraint equation $\partial^2 X_0^I=0$.
This vanishes at  $p=2$  only.
Similarly, $\tilde{\delta} \Psi_0$ is given by 
\begin{align}
\tilde{\delta} \Psi_0(x) 
 & = -2(3-p)\epsilon \cdot x \Psi_0 -\epsilon_{ \mu \nu \lambda } \epsilon^\nu x^\lambda \Gamma^\mu \Psi_0 -(2\epsilon \cdot x x^\mu-\epsilon x^2) \partial_\mu \Psi_0
\end{align}
and  satisfies
\begin{align}
\Gamma^\alpha \partial_\alpha (\tilde{\delta}\Psi_0(x))
& = 2(p-2)\Gamma^\alpha \epsilon_\alpha \Psi_0
\end{align}
where we used the constraint equation $\Gamma^\alpha \partial_\alpha \Psi_0=0$.
Again
$\Gamma^\alpha \partial_\alpha (\tilde{\delta}\Psi_0(x))=0$
vanishes at  $p=2$ only.
Both of the constraints are compatible with the generalized
conformal transformations at $p=2$.
It shows a specialty of M2 (or D2)  branes. 

We have shown that the constraint equations are compatible
with the generalized conformal transformations.
If the solutions are restricted to  constant ones
as in (\ref{constantsolution}), 
we no longer have the generalized conformal  symmetry. 
It can be maintained only when we consider a  set of 
space-time dependent solutions to the constraint equations.

Recently H. Verlinde \cite{Verlinde:2008di}
also considered space-time dependent solutions
to the constraint equations and discussed the conformal symmetry
of the L-BLG theory. 
In his study the constraint equation is imposed everywhere except at $z_i$ 
where a local operator ${\cal O}_i(z_i)$ is inserted, 
\begin{equation}
X_0^I(x) = \sum \frac{q_i^I}{|x-z_i|} .
\label{insertion}
\end{equation}
This is an inhomogeneous solution to the equation
\begin{align}
\partial^2 X_0^I = - 4 \pi \sum q_i^I \delta^3(x-z_i).
\end{align}
We can add the homogeneous solutions to the above. 
If $q^I$ and  $z$ (omitting the index $i$) transform as
\begin{align}
\delta q^I & = \epsilon \cdot z q^I \notag\\
\delta z_\mu &=2 (\epsilon \cdot z) z_\mu - \epsilon_\mu z^2,
\end{align}
the transformation of $X_0^I$
\begin{equation}
\delta X_0^I(x) = - (\epsilon \cdot x ) X_0^I(x) 
\end{equation}
is reproduced and the L-BLG action is invariant
under the conformal transformations. 
Note that $q^I$ cannot be a constant. 
If $q^I$ is kept fixed, the set of solutions is not closed
under the conformal transformations.
In order to recover the conformal invariance, 
$q^I$  should be a position $z$-dependent charge.

We have shown that the L-BLG theory has 
both of the $SO(8)$ invariance and the conformal
symmetry.
In the next section we discuss the 
symmetry properties of the gravity dual of 
the ABJM theory.

\section{$SO(8)$ and Conformal Symmetry  in Dual Geometry}
\setcounter{equation}{0}
\subsection{Large $k$ limit of  ABJM geometry}
In the paper \cite{Aharony:2008ug}, it was pointed out that
the $U(N) \times U(N)$ ABJM theory is dual to the M-theory on 
$AdS_4 \times S^7/{\bf Z}_k$, which is a $d=11$
supergravity solution of M2 branes probing the orbifold
${\bf C}^4/{\bf Z}_k$.  We first review the solution of 
supersymmetric M2 branes in $d=11$ supergravity.

The $d=11$ metric of the multiple M2-branes is given by
\begin{align}
 &ds^2 = H^{-\frac{2}{3}}  \left(\sum_{\mu,\nu=0}^2
                        \eta_{\mu \nu} dx^\mu  dx^\nu  \right)
 + H^{\frac{1}{3}}\left(dr^2 + r^2 d\Omega_7^2\right),\notag\\
 & H(r) \equiv 1 + \frac{R^6}{r^6},
  \label{M2-brane solution}
\end{align}
where $R^6= 32 \pi^2 N' l_p^6 $ and
 $d\Omega_7^2$ is the metric of a  unit 7-sphere.
$N'$ is the number of the M2 branes and identified with
$N'=kN$.
The three form field is also given as
\begin{equation}
C^{(3)} = H^{-1} dx^0  \wedge dx^1 \wedge dx^2
\end{equation}
and the 4-form flux normalized by the world volume is proportional to $N'$.

By focusing on the near horizon region of the  M2-brane, 
the geometry becomes $AdS_4 \times S^7$ geometry.
In the near horizon limit 
$\  R \gg  r $, 
$H(r)$ is replaced by $H(r)=(R/r)^6$ and the metric 
becomes
\begin{align}
ds^2 &= \left( \frac{r}{R} \right)^4 \left(\sum_{\mu,\nu=0}^2
                        \eta_{\mu \nu} dx^\mu  dx^\nu  \right)
                        + \left( \frac{R}{r} \right)^2 dr^2
                        + R^2 d\Omega_7^2  \notag\\
&= R^2  \left[ 
\frac{1}{4} ds^2_{AdS}  + d\Omega^2_7
\right]                  
\end{align}
where we have rescaled the M2 brane world volume coordinates by a factor
$2/R^3$. 
Hence the  near horizon geometry of the supersymmetric M2  branes is given 
by $AdS_4 \times S^7$ with a radius $R$. 
In the large $N'=kN$ limit, the radius becomes much larger 
than the $d=11$ Planck length and the $d=11$ supergravity
approximation is valid.

The ABJM theory describes M2 branes on ${\bf C}^4/{\bf Z}_k$
orbifold. The dual geometry can be  obtained by 
first specifying the polarization (choice of the complex coordinates)
in ${\bf R}^8$ and then 
dividing ${\bf C}^4$  by ${\bf Z}_k.$

Since $S^7$, parameterized by $z^A$ ($A=1, \cdots ,4$) with
$|z^A|^2=1$, 
is a $U(1)$-fibration on ${\bf CP}^3$,
the metric of $S^7$  is  written as
\begin{align}
 d\Omega_7^2 &= \left(d\varphi' + \omega \right)^2 + 
ds_{{\bf CP}^3}^2
\end{align}
where $\varphi'$ is the overall phase of $z^A$.
The details of the definition of coordinates are written in Appendix \ref{appendixB}.

We now perform the ${\bf Z}_k$ quotient by 
dividing the overall phase of each $z^A$, namely the 
$\varphi'$ direction. 
By rewriting $\varphi' = \varphi/k$ with 
$\varphi \sim \varphi + 2\pi$, the metric of $S^7/{\bf Z}_k$ becomes
\begin{align}
 ds^2_{S^7/{\bf Z}_k} = \frac{1}{k^2}
  \left(d\varphi + k \omega \right)^2 + ds^2_{{\bf CP}^3}.
  \label{S7/Zk}
\end{align}
Before performing the ${\bf Z}_k$ quotient, the metric 
has the conformal symmetry associated with
the $AdS_4$ geometry and $SO(8)$ symmetry of $S^7$.
The orbifolding breaks the $SO(8)$ symmetry to $SU(4) \times U(1)$ but
the conformal invariance  still exists. 
This is the bosonic symmetry of the ABJM theory.

The L-BLG action can be derived by taking the 
scaling limit (\ref{scaling}) of the
ABJM theory. 
In the gravity side, this scaling corresponds to 
locating the probe M2 branes far from the orbifold
singularity and taking the large $k$ limit.
As we show in the next section, 
the former process recovers the $SO(8)$
if the positions of the M2 branes are considered to be 
dynamical variables.
The latter makes the radius of the $\varphi'$ circle 
small and $d=11$ geometry is reduced to $d=10.$

Now we consider the $k \rightarrow \infty$
limit of the dual geometry of the ABJM theory.
Following the prescription of ABJM,
we shall  interprete the coordinate $\varphi$ as the compact
 direction in reducing from M-theory to type IIA superstring.
Using the reduction formula \cite{Witten:1995ex}
\begin{equation}
ds_{11}^2 = e^{-\frac{2}{3}\phi} ds_{10}^2
  + e^{\frac{4}{3}\phi} (l_p)^2 \left(d\varphi + A\right)^2
  \end{equation}
we get the $d=10$  metric and the dilaton field in
type IIA supergravity as
\begin{align}
 ds_{10}^2 &= \frac{r}{k l_p} H^{-\frac{1}{2}}\left(\sum_{\mu,\nu=0}^2
                        \eta_{\mu \nu} dx^\mu  dx^\nu  \right)
 + \frac{r}{k l_p}
 H^{\frac{1}{2}}\left(dr^2 + r^2 ds_{{\bf CP}^3}^2\right) ,
\label{10metric}
\\
 e^{2 \phi} &= \left(\frac{r}{k l_p}\right)^{3} H^{\frac{1}{2}}
 = \left( \frac{R}{k l_p} \right)^3.
  \end{align}
Hence in the $k \rightarrow \infty$ limit,
the metric becomes $AdS_4 \times {\bf CP}^3$:
\begin{equation}
 ds_{10}^2 = \frac{R^3}{k}
  \left[\frac{1}{4}ds_{AdS_4}^2 + ds^2_{{\bf CP}^3}\right]
    \label{general AdS x CP3}
  \end{equation}
 where the radius of curvature in string units is 
\begin{equation}
R_{\rm str}^2 
= \left( \frac{R}{l_s} \right)^2 =
 \frac{R^3}{k l_p^3} = 2^{5/2}\pi \sqrt{\frac{N}{k}}.
\end{equation}
The dilaton is a constant 
and this is the reason why the $d=10$ metric
still has a conformal symmetry associated with the $AdS_4$ geometry. 
This is different from the ordinary reduction of the M2 branes to D2 branes
by compactifying the 11th direction of the Cartesian coordinate
(see Appendix \ref{sec:ordin-reduct-m2}).
Note that in the type IIA picture, in addition to the  four-form  RR flux
$F_4$, there is a 2-form RR flux:
\begin{align}
F_4 &= \frac{3}{8} \frac{R^3}{l_p^3} \hat{\epsilon}_4, \notag\\
F_2 &= dA =k d \omega
\end{align}
where $\hat{\epsilon}_4$ is the volume form in a unit
radius $AdS_4$ space. 
Hence the geometry is described 
by the $AdS_4 \times {\bf CP}^3$ compactification
with $N$ units of the four form flux on $AdS_4$
and $k$ units of the two-form flux on the ${\bf CP}^1$
in ${\bf CP}^3$ space.

In the $k  \rightarrow \infty$ limit with 
$N/k$ fixed, the compactification 
radius along the $\varphi$-direction $R_{11}$
becomes very small compared to the 
$d=11$ Planck length:
\begin{align}
  \frac{R_{11}}{l_p} =\frac{R}{k l_p} &\sim \frac{(N k)^{1/6}}{k}\rightarrow 0.
\end{align}
Thus the theory is reduced to a ten-dimensional type IIA 
superstring on $AdS_4 \times {\bf CP}^3$. 
However the scaling limit from ABJM to L-BLG is
taking large $k$ limit before taking the large $N$
and the 't Hooft coupling $N/k$ becomes $0$
in this limit. Since $R_{11} = g_s^{2/3} l_p$,
the string coupling constant $g_s=e^\phi$ 
also becomes $0$:
 \begin{align}
 g_s= e^\phi &\sim k^{-\frac{5}{4}}N^{\frac{1}{4}} \rightarrow 0.
\end{align}
Since $d=11$  Planck length $l_{p}$
and $d=10$ Planck length $l_p^{(10)}$ 
are related to the string length as $l_p=g_s^{1/3} l_s$
and $l_p^{(10)}= g_s^{1/4} l_s$,
the ratios of the radius of the 
IIA geometry~(\ref{general AdS x CP3}) 
with $l_s$ and $l_p^{(10)}$
are  given by 
\begin{align}
\left( \frac{R}{l_s} \right)^2
 \sim \sqrt{\frac{N}{k}}  \rightarrow 0, 
 \ \ \ \ \ \ 
\left( \frac{R}{l_p^{(10)}} \right)^2 
 \sim k^{1/8} N^{3/8} \rightarrow \infty.
\end{align}
Therefore the Type IIA supergravity approximation itself 
is good but the $\alpha'$ expansion is not good
and the theory cannot be considered as 
the low energy approximation of type IIA superstring.
On the other hand, the radius $R$ is much larger than
the $d=11$ Planck length
and it may be more appropriately   interpreted as a dimensional
reduction of M2 branes in  the $d=11$ supergravity.

We summarize the various length scales in the 
scaling limit of the ABJM theory to the L-BLG theory:
\begin{align}
R_{11} \ll l_p^{(11)} \ll l_p^{(10)} \ll R_{AdS} \ll l_s.
\end{align}
The compactification radius $R_{11}$ is much smaller 
than any other scales and the theory is 
reduced to $d=10$.
But the radius of the $AdS_4 \times {\bf CP}^3$
is smaller than the string length and larger
than the $d=10$ and $d=11$ Planck scales.
 
In the ordinary case of the duality between 
type IIB superstrings on 
$AdS_5 \times S^5$ and ${\cal N}=4$ SYM in $d=4$,
the radius of curvature  $R$ is given by
\begin{align}
\left( \frac{R}{l_s} \right)^4 \sim  g_s N, \ \ \ \ \ 
\left( \frac{R}{l_p^{(10)}} \right)^4 \sim N .
\end{align} 
Thus it is usually assumed that both of 
$g_s N$ and $N$ are large so that the type IIB supergravity
approximation and the $\alpha'$-expansion are
valid. Unless $g_s N$ is large, 
$\alpha'$ corrections cannot be neglected and 
the supergravity description itself is not valid.
In the weak coupling limit, 
the dual field theory is usually considered to be
more appropriate.
In our case, we can consider the 
$d=10$ supergravity as a dimensional reduction
of $d=11$ supergravity.
However membranes wrapping the $\varphi$ direction become
very light strings in the unit of the radius of curvature $R$, 
and this may invalidate the supergravity approximation of the M-theory.
\subsection{Recovery of $SO(8)$ in dual geometry of
 L-BLG\label{sec:recovery-so8-dual}}
In taking the scaling limit $k (\gg N) \rightarrow \infty$
of the ABJM theory to the L-BLG theory, the eleven-dimensional geometry is reduced to
the ten-dimensional  $AdS_4 \times {\bf CP}^3$: 
\begin{align}
ds^2 &= H^{-\frac{2}{3}} 
\left( \sum \eta_{\mu \nu}dx^\mu dx^\nu \right)
+ H^{\frac{1}{3}} (dr^2 + r^2 ds^2_{{\bf CP}^3})
\notag\\
H(r)& =\frac{R^6}{r^6}.
\label{LBLGmetric}
\end{align}

In this section we discuss how the $SO(8)$ can be
recovered in the scaling limit of the ABJM geometry
to the L-BLG geometry. 
The L-BLG geometry is obtained by taking $k \rightarrow \infty$
limit of $AdS_4 \times S^7/{\bf Z}_k$ and
simultaneously locating the probe M2 brane far from 
the origin of the orbifold. 
In the large $k$ limit, the geometry  becomes
$d=10$ $AdS_4 \times {\bf CP}^3$, and 
 there are only 7 transverse directions to the M2 brane world volume,
However the radial distance in (\ref{LBLGmetric}) is 
given by the distance in $d=8$:
\begin{align}
 r^2 = \sum_{I=1}^8 (X^I)^2.
\end{align} 
It is invariant under the original $SO(8)$ rotation and 
the ${\bf Z}_k$ quotient leaves $r$ invariant.

Now we consider a probe M2 brane in the above geometry.
In the static gauge, the M2 brane world volume is
identified with the coordinates $x^\mu$ ($\mu=0,1,2$)
 and the
position of the M2 brane is given by $X^I(x)$
where $I=1, \cdots ,8$.  There are only 7 independent
propagating modes among 8, and
the direction that is removed is the $\varphi$-direction.
Remember that the $\varphi$ is the overall phase
of the complex coordinate $z^i$ of the 
transverse $R^8$.
Assuming that the probe M2 brane is located far from 
the source branes, we can separate the probe M2 brane
coordinates into the classical background fields 
$X_0^I(x)$ and the quantum
fluctuations $\hat{X}^I(x)$.
Since the M2 brane is on ${\bf C}^4/U(1)$,
all the points on the gauge orbit 
generated by the $\varphi$-rotation
are identified.
Here the position of the M2 brane is 
represented by the coordinates of ${\bf R}^8$;
a point on the gauge orbit is singled out by fixing
the gauge (see Appendix \ref{appendixB}).

If the probe M2 brane is located at 
\begin{align}
X_0^I = v \delta^{I,8}
\label{x8}
\end{align}
where $v$ is much larger than the scale of the
fluctuations, the rotation along the 
$\varphi$-direction is approximated by
\begin{align}
\delta X^7 &= -  \delta \varphi \ v, \notag\\
\delta X^I&= 0 \ \ , \ \ I \neq 7.
\label{x8phi}
\end{align}
This shows that in the large $v$ limit the $\varphi$
direction can be identified with
the 7th direction 
$X^7$ \footnote{
(\ref{x8}) has fixed a gauge of the $\varphi$ rotation
and  (\ref{x8phi}) is nothing but the direction
parallel  to the gauge orbit.
If we change a gauge,e.g. to $X_0^I=v \delta^{I,7}$, 
(\ref{x8phi}) is also changed accordingly. }.
 Since the ${\bf Z}_k$
orbifolding with large $k$ corresponds to 
gauging away the $\varphi$-direction,
the fluctuation along the 7th direction
is killed and the field $\hat{X}^I$ can fluctuate
only in the other 7 directions.
This means that the $SO(7)$ rotation acts among
the other 7 directions around the classical 
background of (\ref{x8}).  
If the classical background  $X_0^I(x)$ takes  different
directions at different world volume points,
the killed direction also changes locally on the 
world volume.

In order to get a manifest $SO(8)$
covariant formulation of this mechanism,
it is convenient to 
 separate the classical background field of the M2 brane
and the fluctuations in the complex coordinates as
\begin{align}
Z^A(x) = Z_0^A(x) +  \hat{Z}^A(x).
\label{Zi}
\end{align}
If the fluctuations are much smaller than the 
classical background field,
the $\varphi$ rotation can be approximated as
\begin{align}
\delta Z^A = i \delta \varphi Z_0^A.
\label{kill}
\end{align}
If we write
\begin{align}
Z_0^A &= X_0^{2A-1}+ i X_0^{2A} \notag\\
\hat{Z}^A &= i \hat{X}^{2A-1}  - \hat{X}^{2A},
\label{complexSO8}
\end{align}
where $A=1 \cdots 4$,
the  propagating degrees of freedom
 along the direction (\ref{kill}) are killed and 
the fluctuations are restricted to obey
\begin{eqnarray}
X_0^I \hat{X}^I =0.
\label{condition}
\end{eqnarray}
Note that the decomposition of the complex fields
into the real and the imaginary parts are different
between the classical background $Z_0^A$ and the
fluctuations $\hat{Z}^A$ in (\ref{complexSO8}).
With this definition, if $X_0^I =v \delta^{I,8}$,
the killed direction becomes the 8th direction of 
$\hat{X}^I$.
We can write the fluctuations perpendicular to 
$\hat{X}^I$ in (\ref{condition}) as
\begin{align}
P^I = \left( \delta^{IJ} - \frac{X_0^I X_0^J}{(X_0)^2} \right) \hat{X}_J.
\label{Y}
\end{align}
This $P^I$ automatically satisfies the condition
(\ref{condition})
and 7 degrees of freedom are projected among  the
 8 degrees of freedom. 
Now everything is written in a manifestly $SO(8)$
covariant way.
The $SO(8)$ covariance is recovered because we have
assumed that the fluctuation is much smaller than
the classical background fields of the probe M2 brane.
This assumption is consistent with the scaling limit of the ABJM 
theory to the L-BLG theory.

Note here that the $SO(8)$ rotation changes the gauge
choice of the $\varphi$ rotation and $SO(8)$ is 
mixed with the $U(1)$ gauge transformation.
Also note that because of the different assignments
of $X^I$ to $Z^A$ for $Z_0$ and $\hat{Z}$, the
$SO(8)$ is different from the original $SO(8)$ before
taking the orbifolding. 

The analysis here and in the previous section
shows why the L-BLG theory 
has both of the conformal symmetry and the invariance
under $SO(8)$.  
The compactification direction along the 
$\varphi$ direction is different from 
the ordinary reduction to $d=10$ by compactifying
the 11th transverse direction.
The dilaton becomes constant and the
$AdS_4$ geometry is preserved. This is the 
reason why there is a conformal symmetry in 
the effective field theory of L-BLG.

The $SO(8)$ invariance is more subtle.
In the scaling limit of ABJM to L-BLG, 
we take $k \rightarrow \infty$ limit and simultaneously
locate the probe M2 brane far from the origin of the
orbifold.
Then the killed direction of the
fluctuations 
by $Z_k$ ($k \rightarrow \infty$) orbifolding
is  given by the $SO(8)$ vector  of
the classical background fields $X_0^I$
after specifying the gauge choice,
and defining the projection operator by using
$X_0^I$ the manifest $SO(8)$ covariance is
obtained.

\subsection{Actions of probe branes in $AdS_4 \times {\bf CP}^3$}
In this section we study  possible forms of the
 effective field theory of probe
M2 branes in the background geometry (\ref{LBLGmetric}).
The analysis in the section follows the prescription
of \cite{Banerjee:2008pda} and \cite{Cecotti:2008qs}
that a classical scalar field in the radial direction is interpreted
as the Yang-Mills coupling.
We will study probe M2 branes
in a curved background while flat 11-dimensional background
is used there.

By using the metric of (\ref{LBLGmetric}), the 
generally covariant  kinetic term 
can be written as 
\begin{align}
 S_0= &-\frac{1}{2}\int d^3 x\,
  \sqrt{-\det g} g^{\mu\nu} g_{IJ}
  {\rm tr} [D_\mu X^{I} D_\nu X^{J} ],
\end{align}
where $\mu,\nu =0,1,2$ are the world volume indices and
$I,J=1, \cdots ,8$ are the target space indices,
and $D_{\mu}=\partial_{\mu} - iA_{\mu}$ is the covariant
derivative to assure that $X^I$ lies on ${\bf C}^4/U(1)$
(see Appendix \ref{appendixB}).

Both of the world volume metric $g^{\mu \nu}$
and the target space metric $g_{IJ}$ are functions of
the position of the M2 branes $X^I(x).$ 
A static gauge is taken and 
the world volume metric $g_{\mu \nu}$ 
is given by the induced metric in the
curved space-time (\ref{LBLGmetric}).

This kinetic term can be simplified as follows.
The metric $g_{\mu\nu}$ and $g^{IJ}$ are functions
of the the M2 brane position through $r$.
As we did in the previous section, we separate
the 8 scalar fields $X^I(x)$ of the probe M2 branes into
a classical background and  quantum fluctuations.
If the probe M2 branes are located far from the
origin of the orbifold singularity, 
the position of the M2 branes
is approximated by the value of the classical 
background fields $X_0^I(x)$ and 
$r \sim \sqrt{(X_0^I(x))^2}$.
Inserting the explicit form of the metric,
the kinetic term can be simplified (see Appendix \ref{appendixB}) as
\begin{align}
  S_0& =-\frac{1}{2} \int\, dx^3 \eta^{\mu\nu} \eta_{IJ}
  {\rm tr} [\partial_\mu P^{I} \partial_\nu P^{J}]
\end{align}
where $P^I(x)$ is the projected fluctuating fields (\ref{Y}).
In deriving this action, we have used that the classical background fields 
$X^I_0$ are slowly varying.
Note that all the dependence of $H(r)$ vanishes
and the kinetic term of the fluctuating fields
 does not have the explicit dependence on 
the position of M2 branes.

The position of the M2 branes $X_0^I$ must satisfy the classical
equation of motion on the geometry (\ref{LBLGmetric}).
Because of the cancellation of $H(r)$, it looks like a
free field equation of motion.
But the fields $X_0^I$ are restricted 
to be on the geometry where
the $\varphi$-direction is killed, and they are 
slightly different
from the constraint equation  (\ref{constraints2})
in the L-BLG theory, or that in the scaling limit of the 
$SU(N) \times SU(N)$ ABJM theory.
This is  related to the effect of the 
$U(1)$ gauge field  of the ABJM theory.
We discuss it in Appendix \ref{sec:u1-part-abjm}.

In the rest of this section,  
we dare to generalize the discussion of the kinetic term
of the scalar field to the other possible terms in the 
 the effective action of the probe M2 branes 
in the geometry (\ref{LBLGmetric}). 
First assume that a gauge field is induced on the effective
action of the probe M2 branes and its action is given by the 
ordinary Yang-Mills type. 
Then the general coordinate invariant YM action in the curved
metric (\ref{LBLGmetric}) is given by
\begin{align}
 - \frac{1}{4} \int d^3 x\,
  \sqrt{-\det g} g^{\mu\rho} g^{\nu\sigma}
  {\rm tr} \left[F_{\mu\nu} F_{\rho\sigma}\right]
   = - \frac{1}{4}\int d^3 x\,
   \left(\frac{R}{r}\right)^2\eta^{\mu\rho}\eta^{\nu\sigma}
   {\rm tr} \left[F_{\mu\nu} F_{\rho\sigma}\right].
\end{align}
(Since we are considering the $d=11$ theory, there is no freedom
to multiply a dilaton dependence in the action.) 
In this case, $H(r)$ dependence remains and the effective Yang-Mills coupling
 is given by the following field dependent value:
\begin{align}
 g_{YM}^2(x) = \frac{r^2}{R^2} = \frac{(X_0^I(x))^2}{R^2}.
\end{align}

Similarly if we assume that
the scalar field acquires a quartic potential,
the general coordinate and $SO(8)$ invariance require
its form to be
\begin{align}
  &\frac{1}{4} \int d^3 x\,
  \sqrt{-\det g} g_{IK} g_{JL}
  {\rm tr} [P^{I}, P^{J}][P^{K}, P^{L}]\notag\\
  & =\int d^3 x\, \frac{1}{4} \frac{(X_0^I)^2}{R^2}
  \eta_{IK} \eta_{JL}
  {\rm tr} [P^{I}, P^{J}][P^{K}, P^{L}].
\end{align}
Here $P^I$ are projected scalar fields (\ref{Y}).

Summing up these three terms, we have the following
forms of the effective action:
\begin{align}
  S & =-\frac{1}{2} \int\, dx^3  \left(
  {\rm tr} [\partial_\mu P^{I} \partial^\mu P^{I}] 
   - \frac{1}{4}
   \frac{R^2}{(X_0^I)^2}
      {\rm tr} \left[F_{\mu\nu} F^{\mu\nu} \right]
 +   \frac{1}{4} \frac{(X_0^I)^2}{R^2}
  {\rm tr} [P^{I}, P^{J}]^2 \right) .
\end{align}
Of course there is little justification of the above
analysis but it is amusing to see that 
this is nothing but the bosonic part of
 (\ref{Janus}). 
The analysis might support an interpretation that
 the action of L-BLG is the effective action of the 
probe M2 branes in the geometry of (\ref{LBLGmetric}). 
The $X_0^I$ dependence of the coefficients will be
related to the conformal invariance of the M2 branes.
It will be interesting to constrain possible forms 
of the effective action including fermions, higher
derivative terms, or generic potential terms by the generalized conformal invariance.


\section{Conclusions}
\setcounter{equation}{0}
In this paper, we investigated the conformal symmetries 
and the recovery of $SO(8)$ invariance of 
the Aharony-Bergman-Jafferis-Maldacena (ABJM) theories 
and Lorentzian Bagger-Lambert-Gustavsson (L-BLG)
theories.
The conformal invariance, in particular, the invariance 
under the special conformal transformations 
does hold in the L-BLG theory
 only when we consider a set of   spacetime dependent
solutions to the constraint equations
$\partial^2 X_0^I=0.$ 
The conformal symmetries in the field theories
are consistent with the gravity duals;
$AdS_4 \times S_7/{\bf Z}_k$ geometry for the ABJM theory
and  $AdS_4 \times {\bf CP}^3$ geometry for the L-BLG.

It may sound strange that
the L-BLG has a $SO(8)$ global symmetry
while the dual geometry $AdS_4 \times {\bf CP}^3$
does not have it manifestly.
In order to resolve the problem, we 
investigated the recovery of $SO(8)$
by considering a dual geometry around 
a classical background.
We have  shown how
 the $SO(8)$ covariance
is recovered in the geometry  probed by a
slowly varying M2 brane located far
from the orbifold singularity.

Although the  radius of $AdS_4$ is
larger than the $d=10$ Planck length and the
type IIA supergravity approximation is good,
it is much smaller than the IIA string scale 
and the dual geometry of the 
scaled theory of L-BLG 
cannot be interpreted as
the low energy effective theory of 
type IIA superstring. But the radius is larger
than the $d=11$ Planck length and it can be
considered as  a dimensional reduction of the $d=11$ supergravity
solution.

We have also studied the effective action of 
probe M2 branes in a curved geometry 
that is obtained by taking the scaling limit
of $AdS_4 \times S^7/{\bf Z}_k$. 
It is amusing and also somewhat surprising that
 the position dependent coefficients of the coupling constant
 can be correctly reproduced; $g_{YM}^2$ is proportional to
 a square of the position of the M2 branes.
 In particular, if we assume that the scalar potential 
is quartic, the potential is shown to be multiplied by
a square of the center of mass coordinates of the M2 branes.
This is consistent with the sextic potential which is expected 
for  the effective theory of M2 branes.

Finally we would like to comment on a 
subtlety  related to the $U(1)$ factor
when interpreting the L-BLG theory as a 
$k \rightarrow \infty$ scaling limit of ABJM theory.
The L-BLG can be obtained by taking the scaling limit
of the $SU(N) \times SU(N)$ ABJM theory.
If the gauge group is $U(N) \times U(N)$, 
the classical background $X_0^I$ must obey 
a classical equation of motion 
 restricted on ${\bf C}^4/U(1)$, not on 
the full ${\bf C}^4.$
This constraint is consistent with the dual
geometric picture of the $U(N) \times U(N)$ ABJM theory.
Thus the original L-BLG theory will be necessary
 to be supplemented by an additional constraint
in order to   interpret it as the M2 branes
probing ${\bf C}^4/U(1)$.

\section*{Acknowledgements}
We thank  Y. Hikida, A. Ishibashi and  S. Moriyama for useful discussions.
The work of S. I. was partly supported by Grant-in-Aid
for Scientific Research. 
\appendix
\section{$U(1)$ part in ABJM theory\label{sec:u1-part-abjm}}
\setcounter{equation}{0}

In scaling the ABJM theory to the L-BLG theory, we have mainly concerned
with the $SU(N) \times SU(N)$ gauge theory. 
In this appendix we consider the scaling
limit of the $U(N) \times U(N)$
ABJM theory, especially the effect of the  $U(1)$ part.
For simplicity we consider the bosonic terms only. 
In the presence of the $U(1)$ gauge field, 
the covariant derivative is modified to 
\begin{eqnarray}
 D_{\mu} Y = \hat{D}_\mu \hat{Y} + 2iB_{0\mu} \hat{Y} + i
  \{\hat{B}_{\mu},\hat{Y}\} + \partial_{\mu} Y_0 + 2i\hat{B}_\mu Y_0 +
  2i B_{0\mu} Y_0,
\end{eqnarray}
where $B_{0\mu}$ is the axial combination of the 
$U(1) \times U(1)$ gauge field 
\begin{align}
B_{0 \mu} =\frac{1}{2}(A_\mu^{(L)} - A_\mu^{(R)}).
\end{align}
The gauge field $B_{0 \mu}$ is associated with
the gauge transformation of the complex field
$Y^A \rightarrow e^{i \varphi} Y^A$. Hence 
if the dual geometry is described by 
${\bf C}^4/U(1)$, we need the gauge symmetry
even after the scaling to L-BLG.
Therefore, we do not scale the $B_{0 \mu}$ field
unlike $B_\mu$.
The scaling is given by 
\begin{eqnarray}
 \hat{B}_\mu \rightarrow \lambda \hat{B}_\mu, \ \ \ \ \ Y_0 \rightarrow
  \lambda^{-1} Y_0, \ \ \ \ \ B_{0 \mu} \rightarrow B_{0 \mu}
\end{eqnarray}
and take the limit $\lambda \rightarrow 0$.
The kinetic term of the scalar fields 
becomes 
\begin{align}
 -\frac{1}{2} \textrm{tr} |D_\mu Y_A|^2 &=
  \textrm{tr}\left[-\frac{1}{2} (\hat{D}_\mu \hat{Y}_A + 2i\hat{B}_\mu
  Y_{0A} + 2i B_{0\mu} \hat{Y}_A)^{\dagger} (\hat{D}^\mu \hat{Y}^A +
 2i\hat{B}^\mu Y_{0}^A + 2i B_{0}^{\mu} \hat{Y}^A)
  \right.\nonumber \\
 & \ \ \ \
  - \frac{(\partial_\mu Y_{0A}+2iB_{0\mu}Y_{0A})^{\dagger}
  (\partial^\mu Y^A_0 + 2i B_0^\mu Y^A_0)}{2\lambda^2} \nonumber \\
 & \left.\ \ \ \
 -i (\partial_\mu Y_{0A} + 2iB_{0\mu} Y_{0A})^{\dagger}\hat{B}^\mu
 \hat{Y}^A
 +i(\partial_\mu Y^A_0 + 2iB_{0\mu} Y^A_0)\hat{B}^{\mu} \hat{Y}^\dagger_A
 \right].
\end{align}
The difference from the $SU(N) \times SU(N)$ case
is that  all the derivative is replaced by the covariant
derivative with respect to $B_{0 \mu}$.
Requiring finiteness of the action,
one can obtain the modified constraint
\begin{align}
D_{U(1)}^2 Y_0^A \equiv (\partial_{\mu} + 
2iB_{0\mu})(\partial^{\mu} + 2iB_0^\mu) Y_0^A = 0.
\end{align} 
The gauge field $B_{0 \mu}$ 
does not have a kinetic term and it
is nothing but the 
auxiliary gauge field $A_\mu$ introduced in the
${\bf C}^4/U(1)$ gauged model discussed in Appendix \ref{appendixB}.

In the presence of the vector-like $U(1)$ gauge field
\begin{align}
A_{0 \mu} = \frac{1}{2} (A_\mu^{(L)} + A_\mu^{(R)}),
\end{align}
there is a coupling of  $B_{0 \mu}$ to $A_{0\mu}$
through the Chern-Simons term.
If we do not scale the $A_{0 \mu}$ either, it is given by
\begin{eqnarray}
 4 \lambda^{-1}K \epsilon^{\mu\nu\rho} \textrm{tr} B_{0\mu} F_{0\nu\rho},
\end{eqnarray}
where  $F_{0\mu\nu} = \partial_{\mu}
 A_{0\nu} - \partial_{\nu} A_{0\mu}$.
Then because of the $\lambda^{-1}$ coefficient
this must vanish too.

If we instead scale the $A_{0\mu}$ gauge field with $\lambda$,
the coefficient becomes of the order $\lambda^0$, and 
an integration over $B_{0 \mu}$ solves it as
\begin{align}
2B_{0\mu}^{(0)} = -\frac{i}{2|Y_0^A|^2} (Y_0^A \partial_{\mu} \bar{\hat{Y}}^A 
- \bar{Y_0}^A \partial_{\mu} \hat{Y}^A) - 
2 K \epsilon_{\mu \nu \rho} F_0^{\nu \rho}.
\end{align}

\section{$SO(8)$ recovery in ${\bf C}^4/U(1)$ model
\label{appendixB}}
\setcounter{equation}{0}
In Section \ref{sec:recovery-so8-dual} 
we showed the recovery of $SO(8)$ invariance
in the scaling limit of $AdS_4 \times {\bf CP}^3$.
In this appendix, we study a ${\bf C}^4/U(1)$ sigma model
and see the  recovery of $SO(8)$. This is a generalization of 
the equivalence of a gauged model on ${\bf CP}^1$  and an $O(3)$ nonlinear $\sigma$ 
model to a higher dimensional target space.

 ${\bf C}^4$ is parameterized by the following angular
variables:
\begin{align}
 z^1 &= \rho e^{i(\phi_1 + \varphi')} \cos\theta,\notag\\
 z^2 &= \rho e^{i(\phi_2 + \varphi')} \sin \theta \cos\psi,\notag\\
 z^3 &= \rho e^{i(\phi_3 + \varphi')} \sin \theta \sin\psi \cos\chi,\notag\\
 z^4 &= \rho e^{i\varphi'}\sin \theta \sin\psi \sin\chi,\notag\\
 0 &\leq \varphi' \leq 2\pi,\quad
 0 \leq \theta, \psi, \chi, \phi_1, \phi_2, \phi_3 \leq \pi.
 \label{complex coordinates}
\end{align}
We then consider a scalar field on ${\bf C}^4/U(1)$
by identifying 
\begin{align}
 z_i \sim e^{i \varphi'} z_i.
\end{align}
The Lagrangian of the scalar field $Z_i(x)$
on ${\bf C}^4/U(1)$
must be invariant under the local gauge transformation
\begin{align}
 Z_i(x) \rightarrow e^{i \varphi'} Z_i(x)
\end{align}
and the action can be written by introducing an auxiliary 
gauge field $A_\mu$ as
\begin{align}
 S=\int d^3 x |(\partial_\mu - i A_\mu)Z_A|^2.
 \label{c4/u1model}
\end{align}
In the ABJM theory, the gauge field comes from
the $U(1)$ part of the
axial combination of the two $U(N)$ gauge fields $B_{0 \mu}$
(see Appendix \ref{sec:u1-part-abjm}). 
The gauge field does not have a kinetic term and 
and it can be eliminated by solving the 
equation of motion as
\begin{align}
A_\mu = \frac{i}{2|Z^A|^2} (Z^A \partial_{\mu} \bar{Z}^A 
- \bar{Z}^A \partial_{\mu} Z^A).
\end{align}
By substituting the solution to the action,
we obtain a nonlinear action which depends on the 
$Z^A$ fields only. 
The action (\ref{c4/u1model}) becomes
\begin{align}
 S = \int d^3 x ( |\partial Z^A|^2 - A_\mu^2 |Z^A|^2 ) .
 \label{NLmodel}
\end{align}
In the case of ${\bf CP}^1$ model, it is well known
that the model is nothing but the nonlinear
$\sigma$-model on $S^2$. 
In our case, it is a nonlinear model on ${\bf C}^4/U(1)$.

Now we expand the field around a classical background
and expand the field as
\begin{align}
 Z^A(x) = Z^A_0 + \hat{Z}^A. 
\end{align}
The classical background satisfies
the equation of motion. 
Assume that the classical background is 
very slowly varying and much
larger than the fluctuation $\hat{Z}^A$:
\begin{align}
|Z^A_0| \gg |\hat{Z}^A| \ , \ |dZ_0^A|.
\label{classicalexpand}
\end{align}
Under the assumption $(\ref{classicalexpand})$,
the quadratic terms of the fluctuations in the action
(\ref{NLmodel}) become
\begin{align}
 S \sim \int d^3 x ( |\partial \hat{Z}^A|^2 - 
A_{\mu}^{(0)2} |Z_0^A|^2 )
\end{align}
where 
\begin{align}
A_{\mu}^{(0)} = \frac{i}{2|Z_0^A|^2} (Z_0^A \partial_{\mu} \bar{\hat{Z}}^A 
- \bar{Z_0}^A \partial_{\mu} \hat{Z}^A).
\end{align}

If we decompose the complex fields into real components
as 
\begin{align}
Z_0^A = X_0^{2A-1} + i X_0^{2A} \notag\\
\hat{Z}^A = i \hat{X}^{2A-1} - \hat{X}^{2A},
\label{complexreal}
\end{align}
the gauge field can be written as
\begin{align}
A^{(0)}_{\mu} = \frac{1}{(X_0^I)^2} X_0^I \partial_{\mu} \hat{X}^I.
\end{align}
Thus the action can be written as a manifestly
$SO(8)$ covariant expression:
\begin{align}
S = \int d^3 x \{ (\partial \hat{X}^I)^2 - 
\frac{1}{X_0^2} (X_0^I \partial \hat{X}^I)^2 \} .
\end{align}
In terms of the projected scalar field
\begin{align}
P^I = \hat{X}^I - \frac{X_0^I X_0^J \hat{X}^J}{(X_0^I)^2},
\end{align}
the action is  written (under the assumption 
(\ref{classicalexpand}))
\begin{align}
S = \int d^3 x (\partial_\mu P^I)^2.
\end{align}

It is manifestly invariant under the $SO(8)$ transformations.
But note that the $SO(8)$ transformation is different 
from the $SO(8)$ acting on the original ${\bf R}^8$
because of the different decompositions of the complex
fields into the real components in (\ref{complexreal}).


\section{Ordinary reduction of M2 to D2\label{sec:ordin-reduct-m2}}
In this appendix, we remind the reader 
of the ordinary reduction  of  M2 branes in $d=11$ supergravity
to D2 branes in $d=10$ type IIA supergravity
to clarify the difference from the reduction 
adopted in the ABJM theory.
By compactifying $x_{11}$ direction and identifying 
$x_{11}\sim x_{11} + 2\pi R_{11}$
the M2 brane solution is given by replacing
the metric 
(\ref{M2-brane solution}) with a  smeared
harmonic function \cite{Aharony:1999ti}
\begin{align}
 H(r) =  \sum_{n=-\infty}^{\infty}
  \frac{R^6}{(r^2 + (x_{11}  + 2\pi n R_{11})^2)^3}.
\end{align}
where $r$ is the radial distance in the 7 non-compact transverse
directions.
The string coupling constant is given by 
$R_{11} = g_s l_s$.
Then we can get the solution of the 
multiple D2-branes  in the string frame
by using the reduction
rule and the 
Poisson resummation at distance much larger than $R_{11}$:
\begin{align}
 ds_{D2} &= H^{-\frac{1}{2}} \left(\sum_{\mu,\nu=0}^2
                        \eta_{\mu \nu} dx^\mu  dx^\nu  \right)
 + H^{\frac{1}{2}}\left(dr^2 + d\Omega_6^2\right),\notag\\
 e^\phi &= H ^{\frac{1}{4}},\notag\\
 H(r) &= \frac{6\pi^2 g_s N l_s^5}{r^5}.
\end{align}
It is quite different from (\ref{general AdS x CP3}). 
Especially the dilaton is not a constant
and the conformal symmetry of the M2 brane geometry
is broken; it is no longer $AdS_4$.
The transverse direction is given by the radial direction and 
$S^6$,  and  therefore  it has the $SO(7)$ invariance. 


\begin{thebibliography}{99}
 \bibitem{Bagger:2007jr}
         J.~Bagger and N.~Lambert,
         ``Gauge Symmetry and Supersymmetry of Multiple M2-Branes,''
         Phys.\ Rev.\  D {\bf 77}, 065008 (2008)
         [arXiv:0711.0955 [hep-th]].
  ``Modeling multiple M2's,''
  Phys.\ Rev.\  D {\bf 75}, 045020 (2007)
  [arXiv:hep-th/0611108].
  ``Comments On Multiple M2-branes,''
  JHEP {\bf 0802}, 105 (2008)
  [arXiv:0712.3738 [hep-th]].


          
 \bibitem{Gustavsson:2007vu}
         A.~Gustavsson,
         ``Algebraic structures on parallel M2-branes,''
         arXiv:0709.1260 [hep-th].
  A.~Gustavsson,
  ``Selfdual strings and loop space Nahm equations,''
  JHEP {\bf 0804}, 083 (2008)
  [arXiv:0802.3456 [hep-th]].
       

 
\bibitem{Gomis:2008uv}
  J.~Gomis, G.~Milanesi and J.~G.~Russo,
  ``Bagger-Lambert Theory for General Lie Algebras,''
  arXiv:0805.1012 [hep-th].

\bibitem{Benvenuti:2008bt}
  S.~Benvenuti, D.~Rodriguez-Gomez, E.~Tonni and H.~Verlinde,
  ``N=8 superconformal gauge theories and M2 branes,''
  arXiv:0805.1087 [hep-th].

\bibitem{Ho:2008ei}
  P.~M.~Ho, Y.~Imamura and Y.~Matsuo,
  ``M2 to D2 revisited,''
  arXiv:0805.1202 [hep-th].

\bibitem{Aharony:2008ug}
  O.~Aharony, O.~Bergman, D.~L.~Jafferis and J.~Maldacena,
  ``N=6 superconformal Chern-Simons-matter theories, M2-branes and their
  gravity duals,''
  arXiv:0806.1218 [hep-th].

        
\bibitem{Bandres:2008kj}
  M.~A.~Bandres, A.~E.~Lipstein and J.~H.~Schwarz,
  ``Ghost-Free Superconformal Action for Multiple M2-Branes,''
  arXiv:0806.0054 [hep-th].
        
\bibitem{Gomis:2008be}
  J.~Gomis, D.~Rodriguez-Gomez, M.~Van Raamsdonk and H.~Verlinde,
  ``The Superconformal Gauge Theory on M2-Branes,''
  arXiv:0806.0738 [hep-th].


\bibitem{Mukhi:2008ux}
  S.~Mukhi and C.~Papageorgakis,
  ``M2 to D2,''
  JHEP {\bf 0805}, 085 (2008)
  [arXiv:0803.3218 [hep-th]].
     
\bibitem{Honma:2008un}
  Y.~Honma, S.~Iso, Y.~Sumitomo and S.~Zhang,
  ``Janus field theories from multiple M2 branes,''
  arXiv:0805.1895 [hep-th].
        
\bibitem{Ezhuthachan:2008ch}
  B.~Ezhuthachan, S.~Mukhi and C.~Papageorgakis,
  ``D2 to D2,''
  arXiv:0806.1639 [hep-th].
      
\bibitem{Honma:2008jd}
  Y.~Honma, S.~Iso, Y.~Sumitomo and S.~Zhang,
  ``Scaling limit of N=6 superconformal Chern-Simons theories and Lorentzian
  Bagger-Lambert theories,''
  arXiv:0806.3498 [hep-th].

\bibitem{Verlinde:2008di}
  H.~Verlinde,
  ``D2 or M2? A Note on Membrane Scattering,''
  arXiv:0807.2121 [hep-th].

\bibitem{Gomis:2008cv}
  J.~Gomis, A.~J.~Salim and F.~Passerini,
  ``Matrix Theory of Type IIB Plane Wave from Membranes,''
  arXiv:0804.2186 [hep-th].
        
\bibitem{Hosomichi:2008qk}
  K.~Hosomichi, K.~M.~Lee and S.~Lee,
  ``Mass-Deformed Bagger-Lambert Theory and its BPS Objects,''
  arXiv:0804.2519 [hep-th].

\bibitem{Blau:2008bm}
  M.~Blau and M.~O'Loughlin,
  ``Multiple M2-Branes and Plane Waves,''
  arXiv:0806.3253 [hep-th].

\bibitem{Schwarz:2004yj}
  J.~H.~Schwarz,
  ``Superconformal Chern-Simons theories,''
  JHEP {\bf 0411}, 078 (2004)
  [arXiv:hep-th/0411077].
        
\bibitem{Gaiotto:2007qi}
  D.~Gaiotto and X.~Yin,
  ``Notes on superconformal Chern-Simons-matter theories,''
  JHEP {\bf 0708}, 056 (2007)
  [arXiv:0704.3740 [hep-th]].
        
        
\bibitem{Bagger:2008se}
  J.~Bagger and N.~Lambert,
  ``Three-Algebras and N=6 Chern-Simons Gauge Theories,''
  arXiv:0807.0163 [hep-th].
        
         
\bibitem{Terashima:2008sy}
  S.~Terashima,
  ``On M5-branes in N=6 Membrane Action,''
  arXiv:0807.0197 [hep-th].
        
\bibitem{Gomis:2008vc}
  J.~Gomis, D.~Rodriguez-Gomez, M.~Van Raamsdonk and H.~Verlinde,
  ``A Massive Study of M2-brane Proposals,''
  arXiv:0807.1074 [hep-th].
        
\bibitem{Hanaki:2008cu}
  K.~Hanaki and H.~Lin,
  ``M2-M5 Systems in N=6 Chern-Simons Theory,''
  arXiv:0807.2074 [hep-th].
       
\bibitem{Hosomichi:2008jb}
  K.~Hosomichi, K.~M.~Lee, S.~Lee, S.~Lee and J.~Park,
  ``N=5,6 Superconformal Chern-Simons Theories and M2-branes on Orbifolds,''
  arXiv:0806.4977 [hep-th].
        
\bibitem{Schnabl:2008wj}
  M.~Schnabl and Y.~Tachikawa,
  ``Classification of N=6 superconformal theories of ABJM type,''
  arXiv:0807.1102 [hep-th].

        
\bibitem{embedding}
  E.~A.~Bergshoeff, M.~de Roo, O.~Hohm and D.~Roest,
  ``Multiple Membranes from Gauged Supergravity,''
  arXiv:0806.2584 [hep-th].
  E.~A.~Bergshoeff, O.~Hohm, D.~Roest, H.~Samtleben and E.~Sezgin,
  ``The Superconformal Gaugings in Three Dimensions,''
  arXiv:0807.2841 [hep-th].

\bibitem{others}
  A.~Mauri and A.~C.~Petkou,
  arXiv:0806.2270 [hep-th].
%
  J.~Bhattacharya and S.~Minwalla,
  arXiv:0806.3251 [hep-th].
    %
  T.~Nishioka and T.~Takayanagi,
  arXiv:0806.3391 [hep-th].
%
   Y.~Imamura and K.~Kimura,
  arXiv:0806.3727 [hep-th].
   %
  J.~A.~Minahan and K.~Zarembo,
  arXiv:0806.3951 [hep-th].
%
  A.~Armoni and A.~Naqvi,
  arXiv:0806.4068 [hep-th].
  %
  A.~Hanany, N.~Mekareeya and A.~Zaffaroni,
  arXiv:0806.4212 [hep-th].
%
   D.~Gaiotto, S.~Giombi and X.~Yin,
  arXiv:0806.4589 [hep-th].
   %
  C.~Ahn,
  arXiv:0806.4807 [hep-th].
      %
  J.~Bedford and D.~Berman,
  arXiv:0806.4900 [hep-th].
        %
  G.~Arutyunov and S.~Frolov,
  arXiv:0806.4940 [hep-th].
%
    B.~.~j.~Stefanski,
  arXiv:0806.4948 [hep-th].
   %
  G.~Grignani, T.~Harmark and M.~Orselli,
  arXiv:0806.4959 [hep-th].
%
  K.~Hosomichi, K.~M.~Lee, S.~Lee, S.~Lee and J.~Park,
  arXiv:0806.4977 [hep-th].
    %
%
  G.~Grignani, T.~Harmark, M.~Orselli and G.~W.~Semenoff,
  arXiv:0807.0205 [hep-th].
%
  S.~Terashima and F.~Yagi,
  arXiv:0807.0368 [hep-th].
%
  N.~Gromov and P.~Vieira,
  arXiv:0807.0437 [hep-th].
%
  C.~Ahn and P.~Bozhilov,
  arXiv:0807.0566 [hep-th].
%
  N.~Gromov and P.~Vieira,
  arXiv:0807.0777 [hep-th].
%
  C.~S.~Chu, P.~M.~Ho, Y.~Matsuo and S.~Shiba,
  arXiv:0807.0812 [hep-th].
%
  S.~Cherkis and C.~Samann,
  arXiv:0807.0808 [hep-th].
   %
  B.~Chen and J.~B.~Wu,
  arXiv:0807.0802 [hep-th].
        %
  Y.~Zhou,
  arXiv:0807.0890 [hep-th].
%
  T.~Li, Y.~Liu and D.~Xie,
  arXiv:0807.1183 [hep-th].
%
  N.~Kim,
  arXiv:0807.1349 [hep-th].
       %
  Y.~Pang and T.~Wang,
  arXiv:0807.1444 [hep-th].
%
  M.~R.~Garousi, A.~Ghodsi and M.~Khosravi,
  arXiv:0807.1478 [hep-th].
%
  A.~Hashimoto and P.~Ouyang,
  arXiv:0807.1500 [hep-th].
%
  D.~Astolfi, V.~G.~M.~Puletti, G.~Grignani, T.~Harmark and M.~Orselli,
  arXiv:0807.1527 [hep-th].
%
    C.~Ahn and R.~I.~Nepomechie,
  arXiv:0807.1924 [hep-th].
%
  D.~Bak and S.~J.~Rey,
  arXiv:0807.2063 [hep-th].
%
  Y.~Imamura and K.~Kimura,
  arXiv:0807.2144 [hep-th].
    %
  B.~H.~Lee, K.~L.~Panigrahi and C.~Park,
  arXiv:0807.2559 [hep-th].
%
    C.~Ahn, P.~Bozhilov and R.~C.~Rashkov,
  arXiv:0807.3134 [hep-th].
       
        
\bibitem{Inonu:1953sp}
  E.~In\"on\"u and E.~P.~Wigner,
  ``On the Contraction of groups and their representations,''
  Proc.\ Nat.\ Acad.\ Sci.\  {\bf 39} (1953) 510.

\bibitem{Jevicki}
  A.~Jevicki, Y.~Kazama and T.~Yoneya,
  ``Generalized conformal symmetry in D-brane matrix models,''
  Phys.\ Rev.\  D {\bf 59}, 066001 (1999)
  [arXiv:hep-th/9810146].

\bibitem{Bandres:2008ry}
  M.~A.~Bandres, A.~E.~Lipstein and J.~H.~Schwarz,
  ``Studies of the ABJM Theory in a Formulation with Manifest SU(4)
  R-Symmetry,''
  arXiv:0807.0880 [hep-th].

        
\bibitem{Benna:2008zy}
  M.~Benna, I.~Klebanov, T.~Klose and M.~Smedback,
  ``Superconformal Chern-Simons Theories and $AdS_4/CFT_3$ Correspondence,''
  arXiv:0806.1519 [hep-th].

\bibitem{Bandres:2008vf}
  M.~A.~Bandres, A.~E.~Lipstein and J.~H.~Schwarz,
  ``N = 8 Superconformal Chern--Simons Theories,''
  JHEP {\bf 0805}, 025 (2008)
  [arXiv:0803.3242 [hep-th]].
        
\bibitem{Witten:1995ex}
  E.~Witten,
  ``String theory dynamics in various dimensions,''
  Nucl.\ Phys.\  B {\bf 443}, 85 (1995)
  [arXiv:hep-th/9503124].
        
\bibitem{Banerjee:2008pda}
  S.~Banerjee and A.~Sen,
  ``Interpreting the M2-brane Action,''
  arXiv:0805.3930 [hep-th].

\bibitem{Cecotti:2008qs}
  S.~Cecotti and A.~Sen,
  ``Coulomb Branch of the Lorentzian Three Algebra Theory,''
  arXiv:0806.1990 [hep-th].
       

\bibitem{Aharony:1999ti}
  O.~Aharony, S.~S.~Gubser, J.~M.~Maldacena, H.~Ooguri and Y.~Oz,
  ``Large N field theories, string theory and gravity,''
  Phys.\ Rept.\  {\bf 323}, 183 (2000)
  [arXiv:hep-th/9905111].


\end{thebibliography}
\end{document}